\documentclass[twocolumn,showpacs,preprintnumbers,amsmath,amssymb]{revtex4}

\usepackage{graphicx}% Include figure files
\usepackage{dcolumn}% Align table columns on decimal point
\usepackage{bm}% bold math
\usepackage{SIunits}

\begin{document}

%\preprint{APS/123-QED}

\title{Bimodal Counting Statistics in Single Electron Tunneling through a Quantum Dot}% Force line breaks with \\

\author{C. Fricke$^{1}$, F. Hohls$^{1}$, W. Wegscheider$^{2}$, and R. J. Haug$^{1}$}
 \affiliation{$^{1}$Institut f\"ur Festk\"orperphysik, Leibniz Universit\"at Hannover,
  Appelstra\ss{}e 2, D-30167 Hannover, Germany \\$^{2}$Angewandte und
Experimentelle Physik, Universit\"at Regensburg, D-93040
Regensburg, Germany}

\date{\today}% It is always \today, today,

\begin{abstract}

We explore the full counting statistics of single electron
tunneling through a quantum dot using a quantum point contact as
non-invasive high bandwidth charge detector. The distribution of
counted tunneling events is measured as a function of gate and
source-drain-voltage for several consecutive electron numbers on
the quantum dot. For certain configurations we observe
super-Poissonian statistics for bias voltages at which excited
states become accessible. The associated counting distributions
interestingly show a bimodal characteristic. Analyzing the time
dependence of the number of electron counts we relate this to a
slow switching between different electron configurations on the
quantum dot.

\end{abstract}

\pacs{72.70.+m, 73.23.Hk, 73.63.Kv}%73.63.Kv, 73.23.Hk, 72.20.My}% PACS, the Physics and Astronomy   73.63.Kv QDs, 73.23.Hk Coulomb blockade; single-electron tunneling 72.20.My Galvanomagnetic and other magnetotransport effects
\maketitle

%%%%%%%%%%%%%%%%%%%%%%%%%%%%%%%%%%%%%%%%%%%%%%%%%%%%%%%% EINLEITUNG %%%%%%%%%%%%%%%%%%%%%%%%%%%%%%%%%%%%%%%%%%%%%%%%%%%%%%%%%%%%%%%%

The dynamics of electron transport through a quantum dot cannot be
accessed by measurements of the average (DC) current alone.
Additional information was successfully gained from measurements
of the shot noise \cite{Blanter_Buettiker, Nauen-2002, Nauen-2004}
and recently even of the 3rd moment \cite{Reulet-2003} of the
current through the quantum dot. But it is hard to see how such
measurements could be extended to even higher moments.

Recently an alternative approach was pointed out for semiconductor
quantum dots: Beside a direct measurement of the resonant
tunneling current through the dot one can also measure the charge
on the quantum dot using a nearby quantum point contact as a
non-invasive and highly sensitive detector \cite{Field-93,
elzerman_prb-03,schleser-04, umladungen}. For sufficiently low
tunneling rates and high detector bandwidth this allows to resolve
individual tunneling events onto and off the dot
\cite{Elzerman_nature-04, Fujisawa-science} and to measure the
full counting statistics  for tunneling through a quantum dot
ground-state \cite{Gustavsson-PRL, Levitov-96}.

We have implemented a measurement of single electron counting with
large bandwidth. This enables us to measure the full counting
statistics for single electron transport through a quantum dot as
a function of both bias and gate voltage for a series of
consecutive electron numbers on the dot. We observe well
understood counting distributions for tunneling through the
ground-states of the quantum dot. But interestingly our analysis
reveals also the occurrence of bimodal counting distributions for
certain numbers of electrons on the dot and sufficient bias
voltage. We relate this to a slow switching between two different
quantum dot configurations that have distinct tunnel couplings to
the leads.

%%%%%%%%%%%%%%%%%%%%%%%%%%%%%%%%%%%%%%%%%%%%%%%%%%%%%%%%%%%%%%%%%%%%%%%%%%%%%%%%%%%%%%%%%%%%%%%%%%%%%%%%%%%%%%%%%%%%%%%%%%%%%%%%%%%%%

Our device is based on a GaAs/AlGaAs heterostructure containing a
two-dimensional electron system (2DES) 34 nm below the surface.
The electron density is $n = 4.59 \cdot 10^{15}
\hspace{1.2mm}\mathrm{m}^{-2}$, the mobility is $\mu = 64.3
\hspace{1.2mm} \mathrm{m^2/V s}$. We have used an atomic force
microscope (AFM) to define the quantum dot (QD) and the quantum
point contact (QPC) structure by local anodic oxidation (LAO) on
the surface \cite{held-98, ullik-00}; the 2DES below the oxidized
surface is depleted and insulating areas can be written.

\begin{figure}[t]
\includegraphics[scale=0.6]{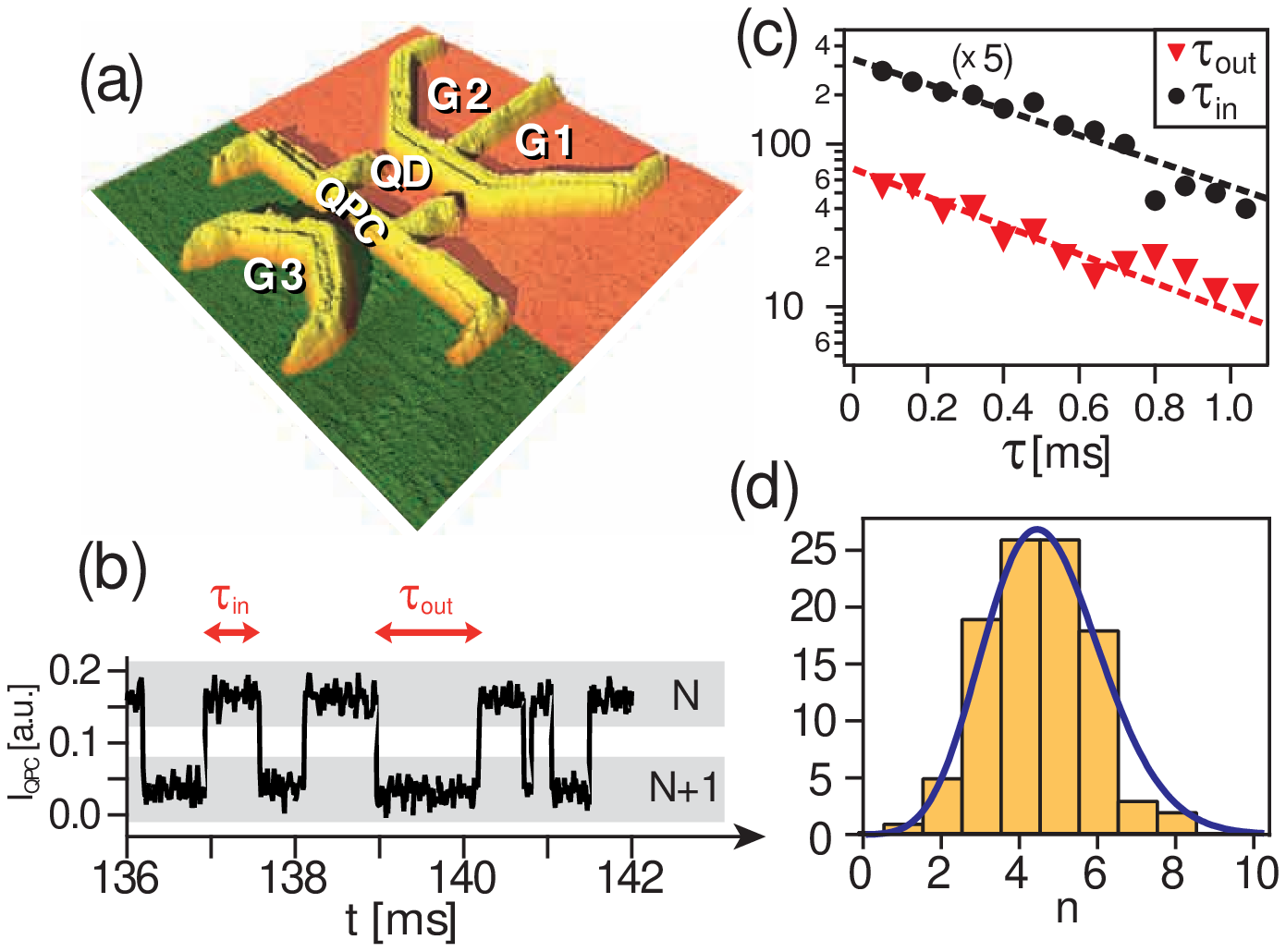}% Here is how to import EPS art
\caption{\label{fig:prinzip} Operating principle of the device
containing a QD and a QPC. (a) AFM image of the device and gate
configuration. (b) Time segment of the QPC signal, tunneling times
$\tau_\mathrm{in}$ and $\tau_\mathrm{out}$ are marked. (c)
Distribution of tunneling times extracted from 100 time segments.
Each time segment spans 5 ms. The distribution for
$\tau_\mathrm{in}$ is shifted by a factor of five for clarity. (d)
Distribution of tunneling events for same measurement compared to
a theoretical model calculation. In total 446 electrons passed the
QD during the measurement leading to this distribution.}
\end{figure}

An AFM image of our device is presented in
Fig.~\ref{fig:prinzip}a. The bright walls depict the insulating
lines written by the AFM. The QPC (left area) is separated from
the QD structure (right area) by an insulating line. The QPC can
be tuned using the in-plane gate G3. The QD is coupled to source
and drain via two tunneling barriers, which can be separately
controlled with gates G1 and G2. These gates are also used to set
the number of electrons in the QD. We use two electrically
separated circuits for simultaneous conductance measurements
through the QPC and the QD. The experimental setup allows us to
resolve tunneling times $\tau$ as small as 30 \micro s. In
Fig.~\ref{fig:prinzip}b the time resolved current through the QPC
is shown. Distinct changes in the current occur for every change
of the number of electrons on the nearby QD. For example the
current drops down whenever an electron enters the dot and thus
the number of electron changes from N to N+1. Due to the
sufficient large bias voltage of 0.2 mV electrons are entering the
QD solely from the source and are leaving the QD to the drain. In
the time segment shown here five traversing electrons can be
identified. Additionally the tunneling times $\tau_\mathrm{in}$
and $\tau_\mathrm{out}$ can be extracted from the QPC signal where
$\tau_\mathrm{in}$ is the time till an additional electron hops
from source onto the dot and $\tau_\mathrm{out}$ is the time that
an electron needs to leave the dot into the drain.

The statistical distributions of the tunneling times
$\tau_\mathrm{in}$ and $\tau_\mathrm{out}$ are shown in
Fig.~\ref{fig:prinzip}c. They follow an exponential decay, where
the exponent is given by the tunneling rates $\Gamma_\mathrm{in}$
and $\Gamma_\mathrm{out}$ respectively. We find
$\Gamma_\mathrm{in} =$ 1.7 kHz and $\Gamma_\mathrm{out} =$ 1.9
kHz. Alternatively one can also determine these rates directly
from the mean value of the tunneling times using the relation
$\Gamma_\mathrm{in(out)}=
\langle\tau_\mathrm{in(out)}\rangle^{-1}$. We find a good
agreement between both methods if the rates are small compared to
the bandwidth of our measurement. For fast rates we will miss a
number of short transitions, which affects the mean of the
tunneling times \cite{Naaman}. Here the analysis of the slope of
the distribution yields superior results.

In this paper we will additionally apply a further method to
extract the statistical properties of the single electron
transport. The measured long time trace is divided into a large
number of short segments as depicted in Fig.~\ref{fig:prinzip}b
and the number of transitions n from N+1 to N electrons is counted
for each segment. We now determine the statistical distribution of
these counts n as depicted in Fig~\ref{fig:prinzip}d. The
experimental results (bars) compare well to the theoretical model
(line) calculated without free parameters from the mean of
$\tau_\mathrm{in}$ and $\tau_\mathrm{out}$ \cite{Bagrets-03,
Gustavsson-PRL}.  From the distribution one can extract not only
the mean value of $n$ but higher moments like the Fano factor
$\alpha$ which is given by the second moment of the obtained
distribution divided by the mean value $\langle n\rangle$:
\begin{equation}\label{eq:alpha}
\alpha = \frac{\langle(n-\langle n\rangle)^2\rangle}{\langle
n\rangle}
\end{equation}

\begin{figure}[tb]
\includegraphics[scale=0.46]{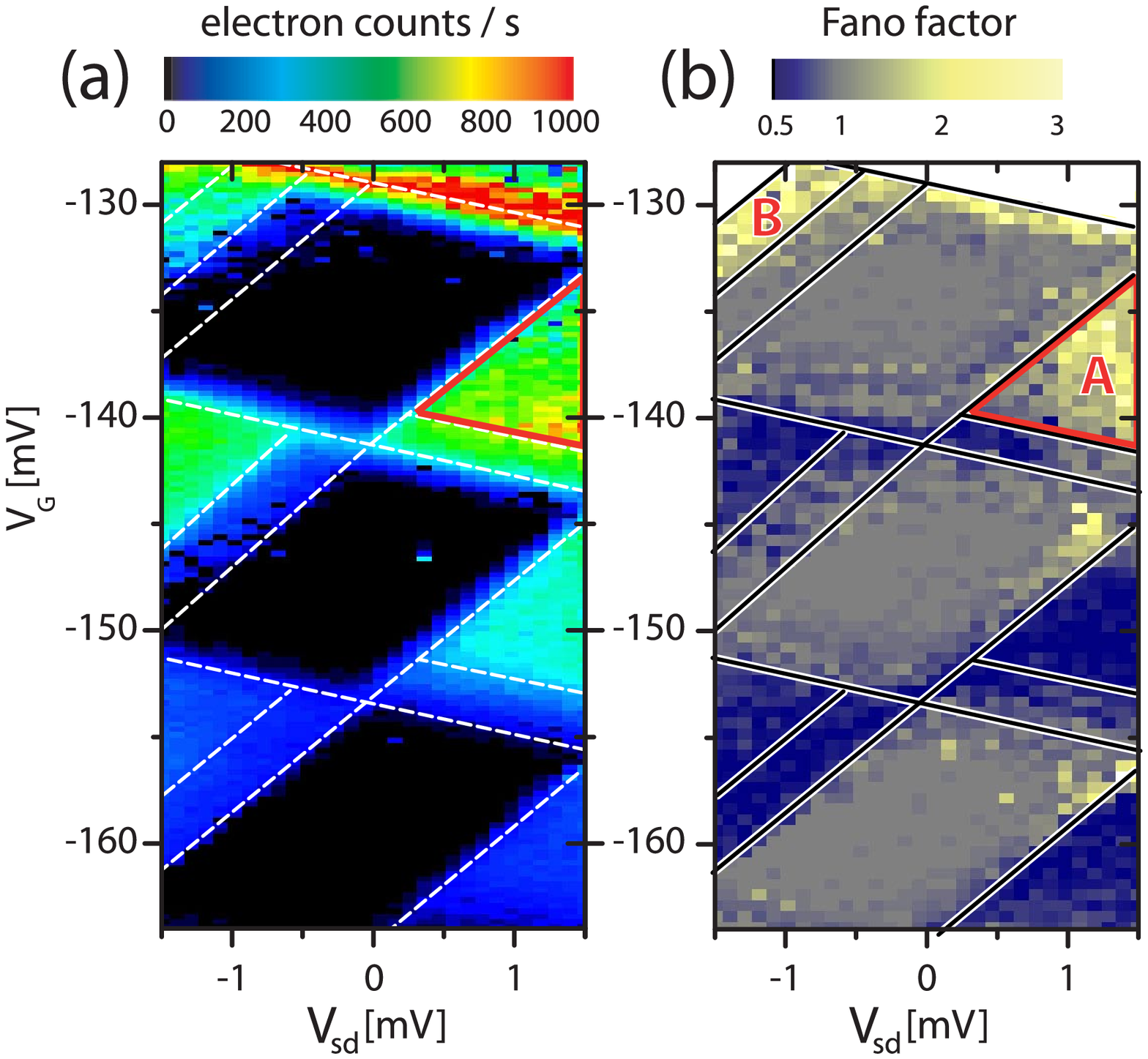}% Here is how to import EPS art
\caption{\label{fig:uebersicht} (a) Counting rate of electrons
passing the QD as a function of gate voltage and source-drain
voltage. The lines mark changes of the average QD occupation which
are deduced from the DC-average of the detector signal. (b) Fano
factor $\alpha$ extracted from the full counting statistics for
the same range. For the white areas above -130 mV at least one of
the tunneling rates is getting to high to extract a reliable
counting statistic.}
\end{figure}

Now we will study the statistics in detail for nonlinear transport
through the QD. In Fig.~\ref{fig:uebersicht}a the rate of
electrons per second traversing the QD as deduced from $\langle
n\rangle$ is shown as a function of gate voltage and bias. Each
data-point represents a full measurement of the distribution of
counting events and the rate is determined from the mean value
$\langle n\rangle$. We clearly observe the so-called Coulomb
diamonds well known from conventional transport experiments. Clear
Coulomb blockade regions are found as well as discrete regions of
finite current due to single electron transport through the ground
state and for large bias also through excited states of the QD.
The QPC detector can also be used to determine the mean charge of
the QD. \cite{maxsymmetrie} For this we analyzed the DC current
through the QPC and extracted the mean charge information. The
changes in the mean charge of the QD are shown as lines in
Fig.~\ref{fig:uebersicht}). The DC charge detection compares well
with the results of the real-time measurement. When the mean
charge of the dot changes a distinct step in the counting rate can
be seen.

In Fig.~\ref{fig:uebersicht}b the Fano factor $\alpha$ as
determined from Eq. \ref{eq:alpha} is shown for the same
measurement. It turns out that $\alpha$ is between $0.5$ and $1$
for most of the QD configurations as one would expect for single
electron transport through the ground state of a QD
\cite{Nauen-2004}. But for a special area marked by the red
triangle in Fig.~\ref{fig:uebersicht} (region \textbf{A}),
super-Poissonian noise is observed. In the marked range not only
the ground-state but also excited states take part in transport.
While the ground-state transport below region \textbf{A} shows
sub-Poissonian characteristic, $\alpha$ rises dramatically when an
excited state enters the transport window. A similar behavior can
also be observed for the opposite transport direction at the next
higher electron number (Fig.~\ref{fig:uebersicht}, region
\textbf{B}). Below region \textbf{B} the ground-state transport is
strongly suppressed but when the excited state takes part in the
transport the current rises and an $\alpha
> 1$ is observed.

\begin{figure}[t]
\includegraphics[scale=0.9]{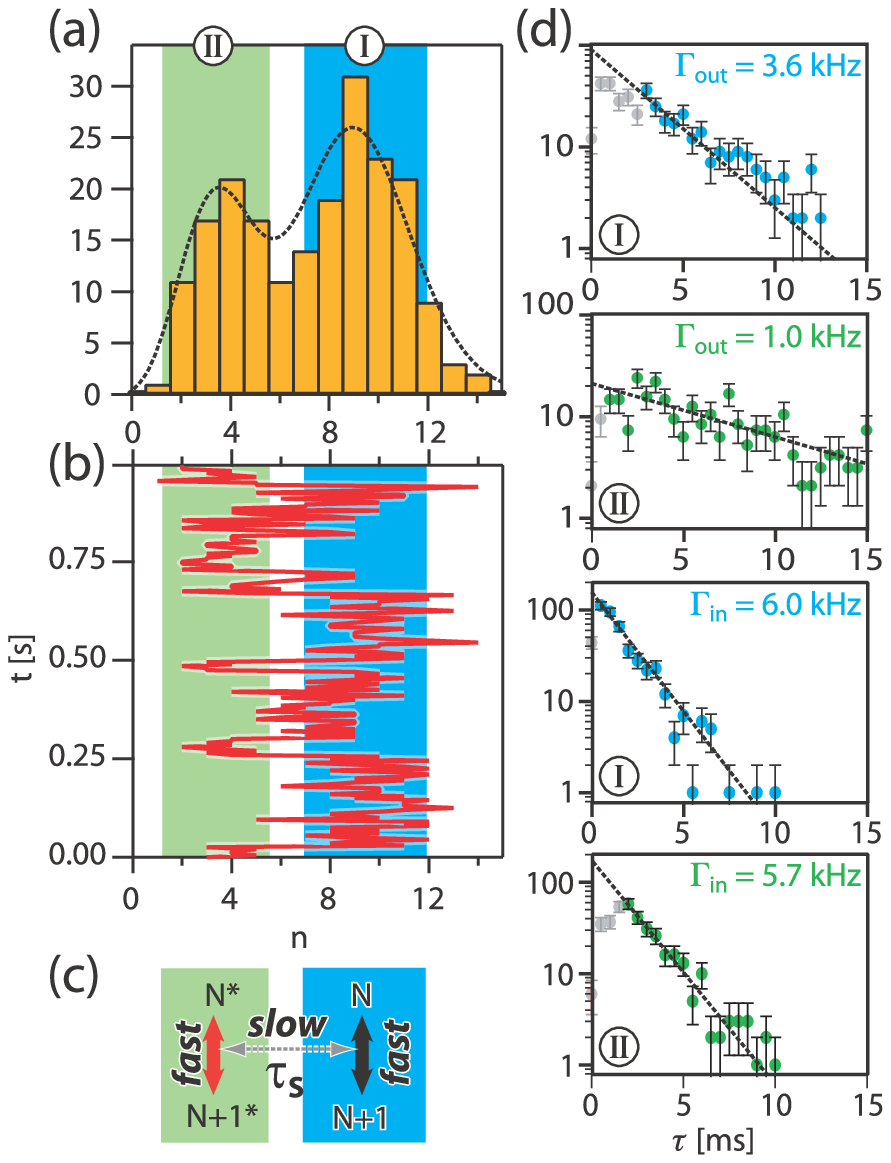}
\caption{\label{fig:bimodal} (a) Typical bimodal distribution of
electrons passing through the dot for the area marked by a
triangle in Fig.~\ref{fig:uebersicht}. Transport channels are
marked by a number and a different background (color online). The
experimental results (bars) compare well to the basic model
calculation (line). (b) Time evolution of the detected number of
electrons per time segment. A switching between the two transport
channels occurs. (c) Simple scheme of the experimental findings.
(d) Distribution of the tunneling times $\tau_\mathrm{in(out)}$
for both transport channels (dots). Short tunneling times (light
gray) are suppressed due to limited bandwidth. The denoted
tunneling rates $\Gamma_\mathrm{in(out)}$ are extracted from the
exponential fit (line). }
\end{figure}

We now analyze the counting statistics for region \textbf{A}
marked by the triangle. For finite bias but only the ground-state
inside the bias-window (directly below region \textbf{A}), we find
no special behavior. Applying the same procedure as described for
Fig.~\ref{fig:prinzip}c we obtain tunneling rates of
$\Gamma_\mathrm{out}=3.4~\mathrm{kHz}$ and
$\Gamma_\mathrm{in}=2.8~\mathrm{kHz}$ for
$V_\mathrm{sd}=1.48~\mathrm{mV}$ and
$V_\mathrm{G}=-143~\mathrm{mV}$ . But as we cross into region
\textbf{A}, i.e. as an excited state enters the bias window, we
find a completely changed counting statistic with a bimodal
distribution. A characteristic distribution is shown in
Fig.~\ref{fig:bimodal}a. In the distribution two clearly distinct
maxima can be identified instead of the single peak structure of
the usual distribution function. Interestingly one of the maxima
emerges at a lower number of counts than the mean $\langle
n\rangle$ observed for tunneling through the ground-state before
an excited state comes into play. How does the distribution relate
to the rate and the Fano factor displayed in
Fig.~\ref{fig:uebersicht}? Firstly the bimodal structure leads to
a significantly higher width of the distribution that results in
the strong increase of the Fano factor. In contrast the mean value
does not show such a dramatic change (compare
Fig.~\ref{fig:uebersicht}a), whereas the mean value of the bimodal
distribution itself has a fairly low probability to be observed.

To study the origin of the bimodality we now analyze the evolution
of the counted electron events with time. This is shown in
Fig.~\ref{fig:bimodal}b. Depicted is the number of electrons
entering the QD in a given time segment in chronological order. By
this we can examine the temporal evolution of all time segments
contained in the distribution of Fig.~\ref{fig:bimodal}a. As
expected from the bimodal distribution there are two separated
ranges where the majority of events are accumulated. These are
marked by the two darker bands in the back of
Fig.~\ref{fig:bimodal}(a,b). Moreover a switching between these
two ranges can be observed. Typically the system stays in each of
this two ranges for a time $\tau_\mathrm{s}$ of 50 to 250
milliseconds what is fairly long compared to the typical tunneling
times $\langle\tau_\mathrm{in/out}\rangle$ of about 0.2 to 1
milliseconds. For the second region showing super-Poissonian noise
(Fig.~\ref{fig:uebersicht}, region \textbf{B}) we found the same
bimodal characteristic and also a comparable switching behavior.
Therefore this bimodality seems to be a more general feature
observable at different electron numbers.

Our observations can be explained by the existence of two
different QD configurations. Each configuration provides a
transport channel for electrons. This behavior leads to
super-Poissonian noise, but differs from the situation described
in \cite{Belzig-05, Gustavson-2006-PRB} for multilevel quantum
dots.

To acquire a deeper understanding of the ongoing processes we
analyzed the tunneling times of the two transport channels in
detail. For this we individually studied tunneling events that can
be related to one of the two QD configurations.

The resulting distributions of the tunneling times
$\tau_\mathrm{in(out)}$ are shown in Fig.~\ref{fig:bimodal}d
separately for configuration \textbf{I} (1st and 3rd graph) and
\textbf{II} (2nd and 4th graph). All sets of data were fitted by
exponential decay to extract the tunneling rates. While the
tunneling barriers were roughly symmetric for the ground state as
deduced from similar values of $\Gamma_\mathrm{in}$ and
$\Gamma_\mathrm{out}$ (see Fig.\ref{fig:prinzip} and the values
given for $V_\mathrm{sd}=1.48~\mathrm{mV}$), they get clearly
asymmetric for the analyzed situation.

When the first excited state enters the transport window, the
tunneling rate $\Gamma_\mathrm{in}$ has risen rather evenly for
both configuration \textbf{I} (6.0 \kilo Hz) and \textbf{II} (5.7
kHz). Both configurations seem to differ only slightly in the
coupling to the source lead. For the drain lead the result is
different. $\Gamma_\mathrm{out}$ for configuration \textbf{I} (3.6
\kilo Hz) is still comparable to the ground-state. In contrast
configuration \textbf{II} shows a tunneling rate
$\Gamma_\mathrm{out}$ of only 1 kHz that is far less than the
outgoing tunneling rate of the ground-state transport. When the QD
is in configuration \textbf{II} tunneling off the dot is
suppressed as the QD seems to couple less efficiently to the drain
lead.

We can model the strong bimodality of the counting distribution
assuming two independent transport channels. Starting from the
full counting statistics theory \cite{Levitov-96} for single state
transport through a QD we combine two distribution functions to
model our experimental results. The probability to measure a
number \emph{n} of traversing electrons is given by:

\begin{equation}\label{prob}
    \small P(n) = q \int_{-\pi}^{\pi}
    e^{-S_1(\chi)-n\chi} \frac{d\chi}{2\pi} + (1-q)\int_{-\pi}^{\pi}
    e^{-S_2(\xi)-n\xi} \frac{d\xi}{2\pi},
\end{equation}
where $S_1$ and $S_2$ are the generating functions for the
separate transport channels and \emph{q} is the probability to
detect transport through channel \textbf{I}. The generating
function of a single level QD has been calculated for
unidirectional tunneling as observed at sufficient large bias
\cite{Bagrets-03}:

\begin{equation}\label{prob2}
    \small \frac{S(x)}{t_0} =  \left(
    \Gamma_\mathrm{in}+\Gamma_\mathrm{out}-\sqrt{\left(\Gamma_\mathrm{in}-\Gamma_\mathrm{out}\right)^2+4\Gamma_\mathrm{in}\Gamma_\mathrm{out}e^{-ix}}
    \right)
\end{equation}
We use the  tunneling rates extracted individually for the two QD
configurations (Fig.~\ref{fig:bimodal}d) which leaves only $q$ as
remaining free parameter. The resulting distribution function
shown in Fig.~\ref{fig:bimodal}(a, dashed line) was received for
$q=0.51$. The good agreement of model and experimental data
further confirms the idea of a switching between two dot
configurations where the dynamics of each configuration
individually is described by a single pair of tunneling rates
$\Gamma_\mathrm{in}$ and $\Gamma_\mathrm{out}$.

The most likely nature of the different configurations is the
excitation of an electron into a non-equilibrium single particle
state that couples only weakly to source and drain. If the
ground-state transition is given by
$\mathrm{(N)}\rightarrow(N+1)\rightarrow\mathrm{(N)}$ we could
relate this to configuration \textbf{I}. Thus the change of
configuration \textbf{I} to \textbf{II} would involve a process
$\mathrm{(N)}\rightarrow\mathrm{(N+1^*)}\rightarrow\mathrm{(N^*)}$
with (N) and $\mathrm{(N+1)}$ the ground states for N and N+1
electrons and $\mathrm{(N^*)}$ an excited state. If the lowest
lying free single particle state is only weakly coupled, we can
carry a single electron current in the cycle
$\mathrm{(N^*)}\rightarrow(N+1^*)\rightarrow\mathrm{(N^*)}$
(compare Fig.~\ref{fig:bimodal}c) until the less probable
$\mathrm{(N+1^*)}\rightarrow\mathrm{(N)}$ transition reinstates
configuration \textbf{I} (ground-state transition).

The two configurations can arise from two origins: spin or charge.
We find that the bimodality only appears when excited states with
a significant excitation energy of 0.3 meV are accessible. We
assume that the occurrence of two transport channels is caused by
two different charge configurations. A change of the spin
configuration should mainly change the tunneling rate into the dot
due to a change of spin selection rule and should have only small
effect on the outgoing rate. In contrast we observe for both
configurations roughly the same tunneling rate for electrons
entering the QD, while a significant difference occurs in the
outgoing rate. The experimental results therefore favor a charge
type effect.

To conclude, we have shown full counting statistics of single
electron tunneling through a quantum dot using a quantum point
contact as non-invasive high bandwidth charge detector. We observe
super-Poissonian noise for certain QD configurations where excited
states take part in electron transport. For these configurations a
clear bimodality of the electron counting distribution occurs. We
analyzed the bimodal distribution in detail and found a slow
switching behavior in the dot transport. We analyzed the tunneling
times for both configurations and presented a model. The good
agreement of model and experimental results confirms the
presumption of two independent transport channels. The switching
between two independent transport channels can be explained by the
existence of two different QD configurations.

This work has been supported by BMBF in the framework of nanoQUIT.

\end{document}